\documentclass{aa}
\usepackage{graphicx}
\usepackage{txfonts}
\begin{document}
   \title{The origin of GEMS in IDPs as deduced from microstructural evolution of amorphous silicates with annealing}

   \author{C. Davoisne\inst{1}
     \and Z. Djouadi\inst{2}
     \and H. Leroux\inst{1}
     \and L. d'Hendecourt\inst{2}
      \and A. Jones\inst{2}
       \and D. Deboffle\inst{2}
          }

   \offprints{H. Leroux, \email{hugues.leroux@univ-lille1.fr}}

   \institute{Laboratoire de Structure et Propri\'et\'es de l'Etat Solide - UMR 8008, Universit\'e des Sciences et Technologies de Lille, 59655 Villeneuve d'Ascq Cedex, France\\
   \and
   Institut d'Astrophysique Spatiale (IAS), B\^atiment 121, F-91405 Orsay (France); Universit\'e Paris-Sud 11 - CNRS (UMR 8617) \\}

   \date{Received ..., 2005; accepted ..., 2005}

% \abstract{}{}{}{}{}
% 5 {} token are mandatory

  \abstract
   {}
     {We present laboratory studies of the micro-structural evolution of an amorphous ferro-magnesian silicate, of
olivine composition, following thermal annealing under vacuum.}
     {The amorphous silicate was prepared as a thin film on a diamond substrate.
Annealing under vacuum was performed at temperatures ranging from
870 to 1020 K. After annealing the thin films were extracted from
the substrate and analysed by transmission electron microscopy to
infer their microstructural and compositional evolution.}
    {Spheroidal metallic nano-particles (2-50 nm) are found
within the silicate films, which are still amorphous after annealing
at 870 K and partially crystallized into forsterite for annealing up
to 1020 K. We interpret this microstructure in terms of a reduction
of the initial amorphous silicate FeO component, because of the
carbon-rich partial pressure in the furnace due to pumping
mechanism. Annealing in a controlled oxygen-rich atmosphere confirms
this interpretation. }
    {The observed microstructures closely resemble those of the GEMS
(Glass with Embedded Metal and Sulphides) found in chondritic IDPs
(Interplanetary Dust Particles). Since IDPs contain abundant
carbonaceous matter, a solid-state reduction reaction may have
occurred during heating in the hot inner regions of the proto-solar
disc. Related to this, the presence of forsterite grains grown from
the amorphous precursor material clearly demonstrates that
condensation from gaseous species is not required to  explain the
occurrence of forsterite around young protostars and in comets.
Forsterite grains in these environments can be formed directly in
the solid phase by thermal annealing of amorphous ferro-magnesian
silicates precursor under reducing conditions. Finally, locking iron
as metallic particles within the silicates explains why astronomical
silicates always appear observationally Mg-rich}
{}
\titlerunning{Origin on GEMS in IDPs}
  \authorrunning{C. Davoisne et al.}
   \maketitle

%________________________________________________________________

\section{Introduction}

Interplanetary dust particles (IDPs) are probably the most pristine materials
in the Solar System. The mineralogy of IDPs could therefore
yield critical information about the processes that occurred in the solar
nebula and the early solar system. IDPs are a fine-grained
 mixture of materials. Among them, GEMS (Glass with Embedded Metal and Sulphides)
 are a major component of the anhydrous IDPs (Bradley \cite{bradley94}).
 GEMS are typically a few 100 nm in size and are composed of a silicate glass
 which includes small (typically 10-50 nm-sized) and rounded grains of Fe, Ni
 metal and sulphide. It has been proposed that the amorphous character of
 GEMS originates from pre-accretional irradiation effect (Bradley \cite{bradley94}).
 Some GEMS also contain some crystalline silicate inclusions of predominantly
  forsterite (e.g., Bradley \cite{bradley03}).
  Some fraction of the GEMS is believed to be of interstellar origin.
  Their infrared spectra compare well with the spectra of the silicate
  grains in molecular clouds, comets and Herbig Ae/Be stars
  (Bradley et al. \cite{bradley99}). Some GEMS display oxygen isotopic anomalies,
  indicating a pre-solar origin (Messenger et al.
  \cite{messenger03}; Floss et al. \cite{floss04}). GEMS with solar oxygen isotopic
  compositions could either have been formed in the early Solar System
   or have been homogenized through processing in the interstellar medium
   (Tielens \cite{tielens03}).
   Another interesting feature of GEMS is that their
bulk elemental compositions are frequently enriched in SiO$_2$, as
shown by the relatively low S/Si, Mg/Si, Ca/Si and Fe/Si ratios,
however, the Al/Si ratio is found to be approximately chondritic
(Schramm et al. \cite{schramm89}; Keller \& Messenger
\cite{keller04a}). Finally, it should be mentioned that GEMS are
frequently found enclosed by, or are in close association with,
carbonaceous materials (e.g., Thomas et al. \cite{thomas93}; Bradley
et al. \cite{bradley99}; Keller et al. \cite{keller00}; Keller et
al. \cite{keller04b}). Despite numerous studies devoted to the GEMS
in IDPs there is currently no consensus regarding their origin and
their formation process. Some GEMS appear to have a clear pre-solar
origin, while others have a solar isotopic composition, which could
be the result of homogenization in the interstellar medium or of
formation in the solar system (Messenger et al. \cite{messenger03}).
An interstellar, shock-accelerated dust origin for these GEMS has
recently been proposed (Westphal \& Bradley \cite{westphal04}).
Alternatively it was also proposed that GEMS which contain relict
crystalline grains could be pseudomorphs formed by irradiation
processing of crystals free-floating in space (Bradley \& Dai
\cite{bradley04}). The amorphous state of the GEMS would therefore
appear to favor an interstellar origin. Indeed interstellar
silicates are believed to be largely amorphous (Mathis
\cite{mathis90}; Demyk et al \cite{demyk99}; Kemper et al.
\cite{kemper04}), an observation that can be explained by low-energy
irradiation in supernova-generated, interstellar shock waves (Demyk
et al. \cite{demyk01}; Carrez et al. \cite{carrez02}). Under this
shock wave irradiation hypothesis the amorphous interstellar
silicates grains could have been progressively heated during their
incorporation into, and their transport through, the proto-planetary
discs. In case of extreme heating they would progressively undergo
annealing, solid-state transformation, melting and volatilization.
Due to solar wind and/or turbulence in the proto-stellar disc, a
fraction of the amorphous silicate particles should have escaped the
severe heating processes.

In this paper we examine the possibility that GEMS could have been
formed by the heating of precursor, amorphous, interstellar grains
in the solar nebula. In our experiments, aimed at understanding the
origin of the GEMS, thin (50-100 nm thick) films of amorphous
silicates, of olivine composition, were annealed under vacuum at
870, 970 and 1020 K on a diamond substrate. The microstructure was
investigated using a transmission electron microscopy (TEM) and
X-ray energy dispersive spectroscopy (EDS) set-up attached to the
TEM.

\section{Experiments}

Interstellar silicates are small grains (typically 10-500 nm
diameter) and have a high surface to volume ratio. We therefore
opted to work with thin films of amorphous silicates (50-100 nm
thick) of olivine composition. The starting sample was a San-Carlos
(Arizona) olivine Mg$_{1.8}$Fe$_{0.19}$Ni$_{0.01}$SiO$_4$. The
amorphous films were prepared by a standard electron-beam
evaporation technique. A current was applied through a tungsten
filament in the vacuum chamber until the thermionic emission of
electrons occurs. By means of electromagnets the electrons are
focussed onto the surface of the sample in a tungsten crucible. The
electron density is sufficient to heat the sample, partially melt it
and form a vapour that condensed on the substrate. The substrate
used in our experiment was a 3 mm diameter diamond disc. This
substrate was chosen for several reasons. Firstly, diamond has an IR
signature far from the absorption regions of silicates (i.e. $\sim$
10 and 20 $\mu$m) allowing thus the IR characterization of the thin
silicate films and their comparison with data from the Infrared
Space Observatory or the Spitzer Space Telescope (Djouadi et al.
\cite{djouadi05}). Secondly, diamond substrates have smooth surfaces
without significant roughness. This makes for an easy extraction of
the thin film for TEM characterization. After their synthesis the
samples were placed in a furnace and were annealed under vacuum at
870, 970 K and 1020 K, at low pressure (10$^{-10}$ bar). In order to
test the influence of oxygen fugacity, some experiments were also
conducted within an O$_{2}$ gas circulation system, at low pressure
(10$^{-7}$ bar). After annealing some areas of the film were scraped
from the substrate and deposited on a carbon-coated TEM foil. The
TEM micro-structural observations were carried out with a Philips
CM30 TEM operating at 300 kV. The crystallographic characterization
was achieved using selected area electron diffraction (SAED). The
microscope was also equipped with an X-ray energy dispersive
spectrometer (EDS) for microanalysis.

\section{Results}

After deposition on the diamond substrate, and removal for TEM
characterization, the thin film is completely amorphous. EDS
measurements show that the composition is close to the starting
San-Carlos olivine, but a slight SiO${_2}$ enrichment is noted. For
the annealing experiment at 870 K, the silicate film remains
amorphous, even for a long duration (up to 780 hrs). For annealing
at 970 and 1020 K (respectively for 55 and 3 hrs), partial
crystallization is observed by IR spectroscopic analysis. The
microscopic investigation we present concern only the samples before
their total crystallization. The main characteristic of all the
samples annealed (at 10$^{-10}$ bar and without O$_{2}$ circulation)
is the presence of widespread iron-nickel nano-particles (Fig. 1)
randomly distributed, 2-50 nm in size, for which the compositions
are highly variable, from 3 to 50 $\%$ Ni. The amorphous phase which
encloses the metallic globules is free of Fe. This microstructure
and microanalyses clearly show that iron, initially in form of FeO,
has segregated from the amorphous phase in the form of metallic
precipitates. Despite the presence of metallic precipitates, the
average composition (amorphous silicate + metallic nano-particules)
is found strongly depleted in Fe. A moderate loss of Mg is also
observed.

The films annealed at 970 and 1020 K contain some forsterite
crystals (Fig. 2). Their grain size ranges from 50 to 250 nm. They
are pure forsterite, i.e., they are iron free. The forsterite grain
abundance varies with the analyzed region, with some areas free of
them, while others make up about 50$\%$ of the surface of the thin
film. The forsterite crystal morphologies are dendritic, as shown by
the irregular growth interface in contact with the remaining
amorphous material. The absence of anhedral morphologies is
compatible with annealing at moderate temperatures.

 \begin{figure}
   \centering
   \includegraphics[width=8.5cm]{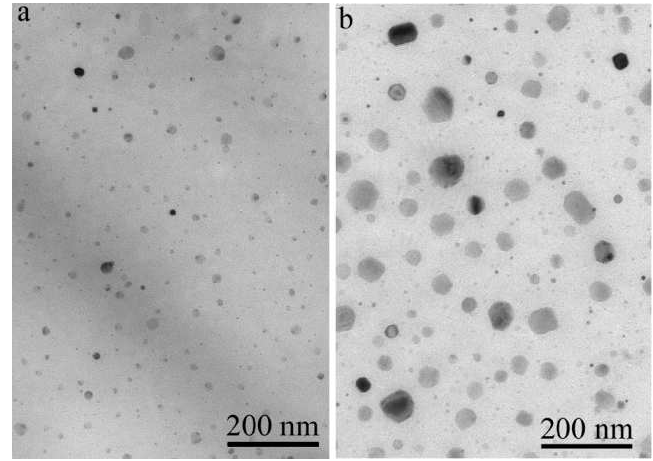}
      \caption{TEM micrograph of annealed sample a) at 870 K for 780 hrs and b) at 1020 K for 3 hrs. Rounded metallic nano-particles enclosed
in the amorphous silicate. They formed by a reduction reaction and
further precipitation since metallic Fe is immiscible in silicates.
The microstructure closely ressembles to those to GEMS found in IDPs
              }
         \label{fig1}
   \end{figure}

 \begin{figure}
   \centering
   \includegraphics[width=8.5cm]{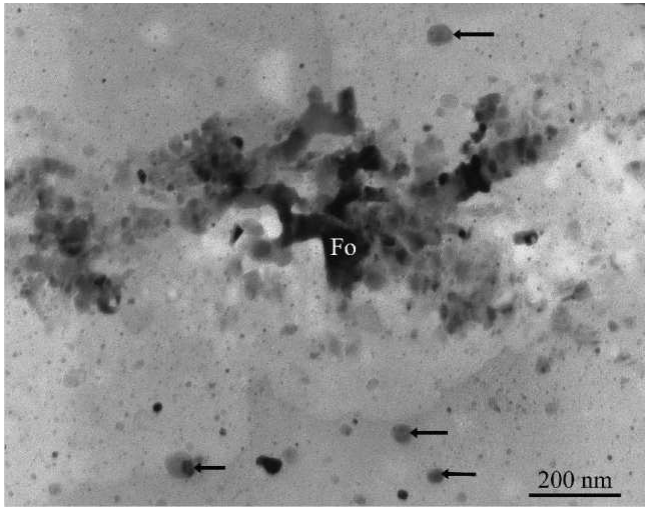}
      \caption{TEM micrograph of sample annealed at 970 K (55 hrs) showing a forsterite crystal (Fo) embedded in a amorphous matrix.
Note the dentritic structure at the edge of the grains. Some metal
particles are also present in the amorphous phase (some of them are
arrowed).
              }
         \label{fig2}
   \end{figure}

Finally, in the samples annealed at higher oxygen pressure
(10$^{-7}$ bar), the amorphous phase of the thin films contain no
metallic precipitates. This comparative experiment shows that the
composition of the residual atmosphere in our vacuum system plays a
crucial role in microstructure development, in particular when the
surface to volume ratio is high as in our thin films.
\section{Discussion}
In this work we have attempted to simulate the incorporation of
amorphous interstellar silicates into the hot inner disc of a
proto-star. We have thus thermally-annealed amorphous
ferro-magnesian silicates under vacuum. The main objective was to
compare the mineralogies obtained in our experiment with those of
the IDPs, in particular the GEMS, some of which are suspected of
having an interstellar origin (Bradley \cite{bradley94}, Bradley et
al. \cite{bradley99}, Messenger et al. \cite{messenger03}). The main
findings of our study are that the annealing of amorphous
ferro-magnesian silicates leads to the formation of forsterite
crystals and metallic iron within a residual amorphous phase. Since
the iron in the starting material was in the form of FeO, the
formation of metallic particles clearly shows that reduction has
occurred according to the reaction: FeO $\rightarrow$ Fe + 1/2
O$_2$. The thermodynamical data associated with this reaction
indicate that the equilibrium oxygen partial pressure is of the
order of 10$^{-20}$ bar at 1000 K (Matas et al. \cite{matas00}). The
total pressure during annealing was 10$^{-10}$ bar. The very low
oxygen partial pressure ($<$ 10$^{-20}$ bar) necessary for the
reduction reaction to proceed is probably due to carbon-rich
contaminants coming  from the pumping system, which consumes oxygen
by the  reaction C + 1/2 O$_2$ $\longleftrightarrow$ CO and thus
induces metal formation according to the reaction FeO + C
$\rightarrow$ Fe + CO. This hypothesis is confirmed by another
annealing experiment performed at higher O$_{2}$ vapor pressure by
injecting an oxygen flux at controlled pressure of 10$^{-7}$ bar. In
this configuration, the carbon-rich species coming from the vacuum
pumps react with the O$_{2}$ gas within the furnace, preventing the
formation of metallic iron on the sample surface.

The presence of iron precipitates within a silicate glass closely
resembles the microstructure observed within the GEMS (e.g., Bradley
\cite{bradley94}). Since IDPs contain a large carbonaceous material
fraction, we suggest that metal formation in GEMS could result from
a reduction reaction during the annealing of amorphous,
ferro-magnesian precursor silicates at temperatures below 1000 K.
This annealing could have occurred within the hot, inner regions of
the proto-solar disc. The occurrence of forsterite crystals in GEMS
is clear (Bradley \cite{bradley03}) and our experiments show that
this occurrence can be explained by an in-situ crystallization of an
amorphous, ferro-magnesian silicate precursor under reducing
conditions. This formation mechanism is also compatible with the
detection of pre-solar (non-solar oxygen isotopic composition)
forsterite crystals in IDPs (Messenger et al. \cite{messenger03})
and presolar olivines and GEMS-like grains in primitive meteorites
(Mostefaoui \& Hoppe \cite{mostefaoui04}). Indeed, the isotopic
composition of the forsterite and olivine grains appears to be close
to that of the precursor materials from which they were formed,
i.e., for some of them presolar amorphous silicate grains.
Forsterite grains are found in cometary grains (e.g., Crovisier et
al. \cite{crovisier97}, Wooden et al. \cite{wooden99}, Wooden et al.
\cite{wooden00}), around young stars (e.g., Waelkens et
al.\cite{waelkens96}, Malfait et al. \cite{malfait98}) and also
around evolved stars (Waters et al. \cite{waters96}), as revealed by
ISO observations. In contrast, interstellar silicates are thought to
be almost completely amorphous (Li \& Draine \cite{li01}, Kemper et
al. \cite{kemper04}), showing that crystalline grains are formed
exclusively within circumstellar environments. Thus, the properties
of the crystalline grains (abundance, chemical composition, grain
size,  etc.) could provide essential information on the conditions
of their formation and their subsequent evolution in discs. For
instance, the crystalline silicates in comets provide an interesting
illustration of this. It has been suggested that crystallization of
the amorphous precursors occurred prior to comet formation, probably
in the hot, inner-edge regions of the accretion disc followed by
radial transport and mixing into the comet-forming region (e.g.,
Bockel\'ee-Morvan et al. \cite{bockelee02}).

In addition to the spectroscopic observations, another source of
information is the study of primitive solar system materials
available for laboratory characterization. For example, the presence
of Mg-rich, crystalline silicates is restricted to very primitive
materials, such as micrometeorites and IDPs (Bradley
\cite{bradley83}, Bradley et al. \cite{bradley92}, Bradley
\cite{bradley03}) and, to date, one meteorite (Mostefaoui \& Hoppe
\cite{mostefaoui04}). Two models are frequently invoked to explain
the presence of crystalline Mg-silicates within circumstellar
environments and in primitive materials. Firstly, they can be formed
via gas-phase condensation. The equilibrium condensation sequence
shows that pure Mg silicates appear first, as predicted in
theoretical models (e.g., Grossman \cite{grossman72}), and as found
in experimental condensation experiments (Rietmeijer et al.
\cite{rietmeijer99}). Another way to form crystalline silicates is
by the thermal annealing of amorphous precursors, such as amorphous,
interstellar silicate grains. In this hypothesis, the formation of
Mg-rich silicates is not obvious, although it has been shown that
iron silicates require higher temperatures to crystallize as
compared to magnesium silicates (e.g., Hallenbeck  et  al.
\cite{hallenbeck00}). Our experimental study shows that forsterite
grains can indeed be formed from amorphous precursors at relatively
moderate temperatures (i.e., around 1000 K) in the presence of
carbon-rich species. This assemblage (silicate+carbonaceous matter)
is typical of the inferred interstellar grain composition and
according to the silicate core-organic refractory mantle model
(e.g., Greenberg \& Li. \cite{greenberg97}). High temperature
gas-phase condensation is thus not exclusively required to explain
the presence of crystalline and magnesium-rich silicates in
circumstellar environments.

\section{Conclusion}
We have performed thermal annealing experiments under vacuum at 870,
970 and 1020 K on an amorphous, ferro-magnesian, silicate thin film
on diamond substrate. The annealed samples were studied by
transmission electron microscopy and energy dispersive spectroscopy.
The main results are the occurrence of spheroidal metallic particles
and the formation of pure forsterite crystals. The overall
microstructures of our annealed samples closely resemble those found
in the GEMS component of the chondritic IDPs. We interpret the
formation of the metallic spheroids, and the accompanying forsterite
crystals, in terms of a reduction of the initial FeO component by
carbon from the reduced atmosphere of our experimental setup. Given
that IDPs contain abundant carbonaceous matter, we propose that
solid-state annealing in association with a reduction reaction at
relatively moderate temperatures (below 1000 K), has occurred during
the heating of amorphous, pre-solar/interstellar, silicate grains in
the hot inner-disc regions of the proto-planetary solar nebula.
Subsequently, this material could have been incorporated into the
outer, cooler and comet-forming regions of the nebula through radial
transport and mixing. Such a process explains also the occurrence of
crystalline Mg-rich silicates observed in comets, the iron being
locked in a metallic nanophase and thus becoming unobservable.

%______________________________________________
\begin{acknowledgements}
The authors would like to thank the French INSU-CNRS "programme
national de plan\'etologie, PNP" for partial financial support of
this work.
\end{acknowledgements}
%-----------------------------------------------------------------

\end{document}